\documentclass[aps,prd,preprint,tightenlines,groupedaddress,showpacs,showkeys]{revtex4-1}

\usepackage{amsmath}
\usepackage{mathrsfs}
\usepackage{hyperref}

\usepackage{feynmp}
\renewcommand{\figurename}{Fig.}
\renewcommand{\tablename}{Tab.}
\newcommand{\rf}[1]{(\ref{#1})}
\newcommand{\rff}[1]{\figurename~\ref{#1}}
\newcommand{\rft}[1]{\tablename~\ref{#1}}
\newcommand{\rfs}[1]{Sec.~\ref{#1}}
\newcommand{\dpr}[1]{\langle#1\rangle}
\newcommand{\cu}{\mathrm{i}}
\newcommand{\rb}[1]{\left(#1\right)}
\newcommand{\bb}[1]{\left[#1\right]}
\newcommand{\df}[2]{\frac{\mathrm{d} #1}{\mathrm{d} #2}}
\newcommand{\eps}{\varepsilon}
\renewcommand{\rho}{\varrho}

\begin{document}
\begin{fmffile}{diagram}

\title{Effective Euler-Heisenberg Lagrangians in models of QED}

\author{Filip P\v{r}eu\v{c}il}
\email{preucil@ipnp.troja.mff.cuni.cz}

\author{Ji\v{r}\'{i} Ho\v{r}ej\v{s}\'{i}}
\email{horejsi@ipnp.troja.mff.cuni.cz}

\affiliation{Institute of Particle and Nuclear Physics,\\
Faculty of Mathematics and Physics, Charles University,\\
V Hole\v{s}ovi\v{c}k\'{a}ch 2, 180 00 Praha 8, Czech Republic}

\begin{abstract}
The main goal of this paper is a direct diagrammatic evaluation of the effective four-photon Lagrangian of the Euler-Heisenberg type for the quantum electrodynamics of massive charged vector bosons. This QED model is naturally embedded in the standard electroweak theory, and we have carried out the corresponding one-loop calculation in the \emph{unitary} gauge. As far as we know, such a work has not been published before since usually the $R$-gauge techniques are preferred for the computation of loop Feynman diagrams. We have recovered the result obtained many years ago by Vanyashin and Terent'ev, who used a specific functional method. For completeness and for validation of our techniques, we have also redone the analogous calculations for scalar and spinor QED.
\end{abstract}

\pacs{12.20.Ds, 14.70.Fm}
\keywords{quantum electrodynamics, effective Lagrangians, light-by-light scattering, massive vector bosons}

\maketitle

\section{Introduction}\label{s:intro}
Effective Lagrangians of the Euler-Heisenberg (EH) type represent a time-honoured topic in quantum field theory (see e.g. \cite{dunne2012} for a historical retrospective). The EH Lagrangian describes a direct interaction of low energy photons (e.g. light-by-light scattering) and represents a quantum correction to the classical Maxwell term. Its evaluation is in fact one of the earliest applications of spinor quantum electrodynamics (QED), as it dates back to the pioneering paper \cite{heisenberg} by Euler and Heisenberg published in 1930s.

A similar work has been done by Weisskopf within the framework of scalar QED \cite{weisskopf}. In modern terms, the results of these seminal papers can be reproduced by means of functional methods used for the calculation of the one-loop effective action for QED in a classical electromagnetic background of constant field strength. Perhaps less known than the classic papers \cite{heisenberg} and \cite{weisskopf} is the work of Vanyashin and Terent'ev \cite{vanyashin}, who have calculated the EH effective Lagrangian within the QED of massive charged vector bosons using also an appropriate functional method. (In fact, the paper \cite{vanyashin} may be better known for providing the first hint of asymptotic freedom in coupling constant renormalization within a non-Abelian gauge theory.) The original results of the above-mentioned papers have been subsequently extended and generalized in various ways --- many relevant results in this area are reviewed in \cite{dunne2005}.

Apart from the aforementioned functional techniques, an alternative way of finding the EH effective Lagrangian consists in a direct evaluation of Feynman diagrams. This amounts to calculating an appropriate one-loop scattering amplitude at the level of the fundamental theory (QED), then performing the low-energy expansion with respect to photon energies and matching the result to the tree-level amplitude corresponding to the form of the EH Lagrangian. Of course, the low-energy expansion in question can be understood equivalently as the large mass expansion with respect to the charged field in the closed loop. The best known ``textbook" example of this sort is the process of light-by-light scattering within spinor QED described by box diagrams made of charged Dirac fields --- the first full evaluation of the relevant one-loop diagrams has been done in \cite{karplus}.

As for scalar QED, the explicit diagram calculations are not so easy to be found in the literature, but in principle one could extract them from some of the existing results concerning the full electroweak standard model (SM) treated within a renormalizable (e.g. 't Hooft-Feynman) gauge. The same could be said about the loops made of charged vector bosons. Some earlier papers to be mentioned in this context are e.g. \cite{konig} and \cite{jiang}, but it should also be noted that matching of the available SM results to the EH effective Lagrangian is by no means straightforward. From the technical point of view, the most challenging task would be a direct calculation of $W$ boson loops within the $U$-gauge in SM. In fact, it is precisely the same as the evaluation of such loops within the QED alone, with the $WW\gamma$ and $WW\gamma\gamma$ couplings of the Yang-Mills type. To the best of our knowledge, such a diagrammatic calculation has not been published previously. Performing such a \emph{tour de force} should bring one an additional bonus of recovering the old result \cite{vanyashin} obtained years ago by means of completely different methods.

The rest of the paper is organized as follows. In the next section, the effective EH Lagrangian is specified, and the corresponding lowest order matrix element for a four-photon scattering process is shown. In \rfs{s:spin}, the familiar case of spinor QED is recapitulated, mostly for reference purposes. The case of scalar QED is treated in some detail in \rfs{s:scal}. \rfs{s:vect} covers the main result of this paper: there the EH effective Lagrangian is evaluated within the QED of massive charged vector bosons. \rfs{s:concl} contains some concluding remarks.

\section{The EH Lagrangian}\label{s:ehl}
As we have noted in \rfs{s:intro}, we will consider here the four-photon effective interactions only (otherwise the computational complexity seems to be prohibitive). To begin with, let us recall the familiar form of the EH effective Lagrangian. Taking into account the electromagnetic gauge invariance as well as the discrete symmetries $C$ and $P$, this can be written as
\begin{equation}
\mathscr{L}_\mathrm{eff.} = g_1(F_{\mu\nu}F^{\mu\nu})^2 + g_2(\star F_{\mu\nu}F^{\mu\nu})^2,
\label{e:lagr}
\end{equation}
where $F_{\mu\nu}$ is the electromagnetic field tensor and $\star F_{\mu\nu}$ its dual, $\star F_{\mu\nu} = \frac{1}{2}\eps_{\mu\nu\rho\sigma}F^{\rho\sigma}$. The numbers $g_1$ and $g_2$ are some effective coupling constants of mass dimension $-4$.

The Feynman rules for the relevant vertices can be extracted from \rf{e:lagr} in a straightforward way, and the lowest order matrix element for a four-photon process can then be written as
\begin{equation}
\mathcal{M} = \Gamma_{\mu\nu\rho\sigma}(p_1, p_2, p_3)\ \eps_1^\mu\eps_2^\nu\eps_3^\rho\eps_4^\sigma,
\label{e:amplit}
\end{equation}
where $p_i$ are the photon four-momenta, and $\eps_i \equiv \eps(p_i, h_i)$ are the corresponding polarization vectors, with $h_i$ denoting the photon helicities. The quantity $\Gamma_{\mu\nu\rho\sigma}$ will be called occasionally the polarization tensor in what follows. The external photons are taken to be on the mass shell, and their physical polarizations are of course transverse. Thus, one has
\begin{align}
p_i^2 &= 0\notag\\
p_i\cdot \eps_i &= 0,\qquad i \in \{1,2,3,4\}.
\label{e:onshell}
\end{align}
The momentum conservation amounts to
\begin{equation}
\sum_{i=1}^4 p_i = 0
\label{e:conserv}
\end{equation}
(we take all the photons as outgoing), and this leads immediately to the identity
\begin{equation}
p_1 \cdot p_2 + p_1 \cdot p_3 + p_2 \cdot p_3 = 0.
\label{e:identity}
\end{equation}
In what follows, we will write the polarization tensor $\Gamma_{\mu\nu\rho\sigma}$ as a function of the three independent momenta $p_1$, $p_2$, and $p_3$ satisfying the constraint \rf{e:identity}, which will be utilized repeatedly in the subsequent considerations. For the sake of brevity of our formulae, we also introduce a shorthand notation
\begin{align}
i_\alpha &\equiv (p_i)_\alpha\notag\\
p_i \cdot p_j &\equiv \dpr{ij},
\end{align}
to be used throughout the paper.

The evaluation of the matrix element \rf{e:amplit} is not difficult but somewhat lengthy, so we will present only the final result. All the necessary calculational details can be found in \cite{preucil}. The polarization tensor $\Gamma_{\mu\nu\rho\sigma}$ can be split naturally into three parts according to the number of fully contracted momenta pairs (i.e. their scalar products). Such a decomposition can be written schematically as
\begin{equation}
\Gamma_{\mu\nu\rho\sigma} = \Gamma_{pppp} + \Gamma_{ppg} + \Gamma_{gg},
\label{e:scheme}
\end{equation}
where the meaning of the used symbols is as follows: $\Gamma_{pppp}$ contains only the terms with the structure $(p_i)_\alpha (p_j)_\beta (p_k)_\gamma (p_l)_\delta$ (this can be written as $i_\alpha j_\beta k_\gamma l_\delta$ in our shorthand notation), $\Gamma_{ppg}$ is composed only of the terms $i_\alpha j_\beta g_{\gamma\delta}\dpr{kl}$, and finally $\Gamma_{gg}$ incorporates only the terms $g_{\alpha\beta}g_{\gamma\delta}\dpr{ij}\dpr{kl}$, where $i,j,k,l \in \{1,2,3\}$ are some external momenta, and $\alpha,\beta,\gamma,\delta\ \in \{\mu,\nu,\rho,\sigma\}$ are some tensor indices.

Before displaying our results for the three terms in \rf{e:scheme}, there are two simple technical points to be mentioned. First, for the on-shell photons, all expressions containing squares of the external momenta $p_1$, $p_2$, or $p_3$ effectively vanish and therefore do not appear in the tensor $\Gamma_{\mu\nu\rho\sigma}$. Further, due to the transversality of the photon polarizations, all longitudinal terms (i.e. those involving at least one element from the set $\{1_\mu, 2_\nu, 3_\rho\}$) are discarded from $\Gamma_{\mu\nu\rho\sigma}$ too.

The final form of the tensor $\Gamma_{\mu\nu\rho\sigma}$ is then the following
\begin{align}
\Gamma_{pppp} &= N \sum_\mathrm{perm.} (g_2 - g_1)1_\nu 1_\rho 2_\sigma 3_\mu - g_2 (1_\rho 1_\sigma 2_\mu 3_\nu)\notag\\
\Gamma_{gg} &= N \sum_\mathrm{perm.} \bb{\rb{\frac{g_1}{2} - g_2} g_{\mu\nu} g_{\rho\sigma} + g_2(g_{\mu\sigma} g_{\nu\rho})}\dpr{12}^2 \notag\\
\Gamma_{ppg} &= N \sum_\mathrm{perm.} \Big[\rb{g_2 - \frac{g_1}{2}} 1_\nu 2_\mu g_{\rho\sigma} + (g_1 - 2g_2) 1_\rho 3_\sigma g_{\mu\nu}\notag\\
&\hphantom{{}= N \sum_\mathrm{perm.} \Big[} + g_2 (1_\rho 3_\nu g_{\mu\sigma} - 1_\rho 2_\mu g_{\nu\sigma} - 1_\nu 3_\mu g_{\rho\sigma}\notag\\
&\hphantom{{}= N \sum_\mathrm{perm.} \Big[ + g_2 (} - 1_\sigma 2_\mu g_{\nu\rho} + 1_\sigma 2_\rho g_{\mu\nu} + 2_\mu 3_\sigma g_{\nu\rho}\notag\\
&\hphantom{{}= N \sum_\mathrm{perm.} \Big[ + g_2 (} + 1_\rho 3_\mu g_{\nu\sigma} - 3_\mu 3_\nu g_{\rho\sigma} + 3_\mu 3_\sigma g_{\nu\rho}) \Big]\dpr{12},
\end{align}
where $N = 32$, and $\sum_\mathrm{perm.}$ denotes a summation over all simultaneous permutations of $\{1, 2, 3\}$ and $\{\mu, \nu, \rho\}$. For example,
\begin{align}
\sum_\mathrm{perm.} 1_\rho 2_\sigma g_{\mu\nu} \dpr{13} = {}&1_\rho 2_\sigma g_{\mu\nu} \dpr{13} + 2_\rho 1_\sigma g_{\nu\mu} \dpr{23} + 3_\mu 2_\sigma g_{\rho\nu} \dpr{31}\notag\\
&+ 1_\nu 3_\sigma g_{\mu\rho} \dpr{12} + 2_\mu 3_\sigma g_{\nu\rho} \dpr{21} + 3_\nu 1_\sigma g_{\rho\mu} \dpr{32}.
\end{align}

The expression for $\Gamma_{ppg}$ can be made a bit shorter if we use the identity \rf{e:identity}. For instance, we can obtain a relation such as
\begin{equation}
\sum_\mathrm{perm.} (1_\rho 2_\mu g_{\nu\sigma} + 1_\nu 3_\mu g_{\rho\sigma} + 1_\rho 3_\nu g_{\mu\sigma}) \dpr{12}= 0
\end{equation}
and many similar ones. Using them, the expression can be shrunk down to just seven terms. For clarity, here is the entire tensor again including the adjusted term
\begin{align}
\Gamma_{pppp} &= N \sum_\mathrm{perm.} (g_2 - g_1)1_\nu 1_\rho 2_\sigma 3_\mu - g_2 (1_\rho 1_\sigma 2_\mu 3_\nu)\notag\\
\Gamma_{gg} &= N \sum_\mathrm{perm.} \bb{\rb{\frac{g_1}{2} - g_2} g_{\mu\nu} g_{\rho\sigma} + g_2(g_{\mu\sigma} g_{\nu\rho})}\dpr{12}^2 \notag\\
\Gamma_{ppg} &= N \sum_\mathrm{perm.} \Big[ g_2 (1_\sigma 3_\mu g_{\nu\rho} - 1_\sigma 3_\nu g_{\mu\rho} - 3_\mu 3_\nu g_{\rho\sigma} + 3_\mu 3_\sigma g_{\nu\rho})\notag\\
&\hphantom{{}= N \sum_\mathrm{perm.} \Big[} +(g_1 - g_2)(1_\rho 3_\mu g_{\nu\sigma} + 1_\rho 3_\sigma g_{\mu\nu})\notag\\
&\hphantom{{}= N \sum_\mathrm{perm.} \Big[} +2g_2 (1_\rho 3_\nu g_{\mu\sigma}) \Big]\dpr{12}.
\label{e:effect}
\end{align}

Clearly, the tensor \rf{e:effect} is symmetric under photon exchange, which correctly reflects the bosonic nature of photons. Also, it is straightforward (though somewhat tedious) to show that the polarization tensor satisfies the Ward identities
\begin{align}
\Gamma_{\mu\nu\rho\sigma} 1^\mu &= 0\notag\\
\Gamma_{\mu\nu\rho\sigma} 2^\nu &= 0\notag\\
\Gamma_{\mu\nu\rho\sigma} 3^\rho &= 0\notag\\
\Gamma_{\mu\nu\rho\sigma} (1^\sigma + 2^\sigma + 3^\sigma) &= 0,
\end{align}
in accordance with the gauge invariance of the EH Lagrangian \rf{e:lagr}. Note that for the direct proof of these identities, the kinematic relations \rf{e:onshell} and \rf{e:identity} are needed.

Finally, we present a formula for the unpolarized differential cross section 
\begin{align}
\overline{\df{\sigma}{\Omega}}(s, \vartheta) &= \frac{s^3(3+\cos^2\vartheta)^2}{16\pi^2}\bb{(g_1-g_2)^2+2(g_1^2+g_2^2)},
\end{align}
which can be obtained by computing the quadratic invariant $\Gamma_{\mu\nu\rho\sigma}\Gamma^{\mu\nu\rho\sigma}$. Note that the momenta 1 and 2 should be reversed (i.e. $1_\mu\to - 1_\mu$, $2_\mu \to -2_\mu$) for this purpose, having in mind that all the photons are taken as outgoing in our original expression. 

As indicated in \rfs{s:intro}, the next step is to compute appropriate one-loop amplitudes of the four-photon process in question within various models of QED, perform the low-energy expansion, and then match the obtained polarization tensors to the effective one. Thus one can determine the corresponding coupling constants $g_1$ and $g_2$. We are going to proceed in this way in the following three sections. Most of the calculations are too long to be presented here explicitly; the interested reader is referred to \cite{preucil} for details.

\section{Spinor QED}\label{s:spin}
Spinor QED is the best known case, and the corresponding result has been described in many textbooks (cf. e.g. \cite{itzykson}). The lowest order contribution to the considered process is given by the familiar one-loop box diagrams of the type shown in \rff{f:box}.
\begin{figure}[ht]
\centering
\begin{fmfgraph*}(115,92)
\fmfleft{i1,i2}
\fmfright{o1,o2}
\fmf{fermion}{v1,v4}
\fmf{fermion}{v2,v1}
\fmf{fermion}{v3,v2}
\fmf{fermion}{v4,v3}
\fmf{wiggly,tension=2}{v1,i1}
\fmf{wiggly,tension=2}{v2,i2}
\fmf{wiggly,tension=2}{v3,o2}
\fmf{wiggly,tension=2}{v4,o1}
\end{fmfgraph*}
\caption{Spinor QED: Only box diagrams contribute. The internal lines represent propagators of the charged massive Dirac field.}
\label{f:box}
\end{figure}
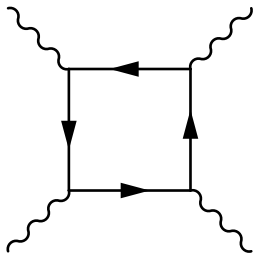

The evaluation of the one-loop amplitude is carried out using standard dimensional regularization technique. It is a well-known fact that while an individual box diagram contains a (logarithmic) ultraviolet (UV) divergence, the full four-photon amplitude is finite since the UV divergences are canceled upon taking into account all relevant permutations of the external photon lines. Along with the UV divergences, some finite terms independent of the external momenta are eliminated as well.

Let us also note that the low-energy expansion of the finite part of the amplitude can be performed before the integration over Feynman parameters, and thus one need not deal with polylogarithms, etc. --- all expressions to be integrated are polynomials. Another crucial consequence of the aforementioned summation over the photon permutations is a cancellation of the finite terms involving only two external momenta: obviously, the survival of such terms would make the envisaged matching of the one-loop amplitude with the form \rf{e:effect} impossible. In other words, these unwanted terms would violate the gauge invariance.

As in the previous section, the polarization tensor is split into three parts
\begin{align}
\Gamma^\mathrm{spin.}_{pppp} &= N_\mathrm{spin.} \sum_\mathrm{perm.} 3(1_\nu 1_\rho 2_\sigma 3_\mu) - 7(1_\rho 1_\sigma 2_\mu 3_\nu)\notag\\
\Gamma^\mathrm{spin.}_{gg} &= N_\mathrm{spin.} \sum_\mathrm{perm.} (3g_{\mu\rho}g_{\nu\sigma} - 7g_{\mu\sigma}g_{\nu\rho})\dpr{12}\dpr{13}\notag\\
\Gamma^\mathrm{spin.}_{ppg} &= N_\mathrm{spin.} \sum_\mathrm{perm.} \Big[ 7(1_\sigma 3_\mu g_{\nu\rho} - 1_\sigma 3_\nu g_{\mu\rho} - 3_\mu 3_\nu g_{\rho\sigma} + 3_\mu 3_\sigma g_{\nu\rho})\notag\\
&\hphantom{{}= N_\mathrm{spin.} \sum_\mathrm{perm.} \Big[} -3(1_\rho 3_\mu g_{\nu\sigma} + 1_\rho 3_\sigma g_{\mu\nu})\notag\\
&\hphantom{{}= N_\mathrm{spin.} \sum_\mathrm{perm.} \Big[} +14(1_\rho 3_\nu g_{\mu\sigma}) \Big]\dpr{12},
\end{align}
where $N_\mathrm{spin.} = 4\alpha^2/45 m^4$, with $m$ being the fermion mass.

One may notice immediately that almost the entire tensor structure in \rf{e:effect} is thus reproduced. The only discrepancy is the $\Gamma^\mathrm{spin.}_{gg}$ term. It is easy to guess that utilizing the identity \rf{e:identity} would now help. Squaring it yields
\begin{equation}
\dpr{12}\dpr{13} = \frac{1}{2}\rb{\dpr{23}^2 - \dpr{12}^2 - \dpr{13}^2},
\end{equation}
which implies
\begin{equation}
\sum_\mathrm{perm.}g_{\mu\rho}g_{\nu\sigma} \dpr{12}\dpr{13} = - \frac{1}{2}\sum_\mathrm{perm.} g_{\mu\nu}g_{\rho\sigma} \dpr{12}^2.
\end{equation}
Substituting this into $\Gamma^\mathrm{spin.}_{gg}$ (the other part is processed in a similar way) then leads to
\begin{equation}
\Gamma^\mathrm{spin.}_{gg} = N_\mathrm{spin.} \sum_\mathrm{perm.} (-5g_{\mu\nu}g_{\rho\sigma} + 7g_{\mu\sigma}g_{\nu\rho})\dpr{12}^2.
\end{equation}
This is precisely what we need for matching with the expression \rf{e:effect}. Thus, we get the familiar effective coupling constants (cf. e.g. \cite{itzykson})
\begin{align}
g_1^\mathrm{spin.} = \frac{4 N_\mathrm{spin.}}{N} &= \frac{\alpha^2}{90 m^4}\notag\\
g_2^\mathrm{spin.} = \frac{7 N_\mathrm{spin.}}{N} &= \frac{7 \alpha^2}{360 m^4}.
\end{align}

\section{Scalar QED}\label{s:scal}
The interaction Lagrangian for scalar QED has the familiar form
\begin{equation}
\mathscr{L}_\mathrm{int.} = -\cu e A_\mu[\phi^\dagger (\partial^\mu\phi) - (\partial^\mu\phi^\dagger)\phi] + e^2 A_\mu A^\mu\phi^\dagger\phi,
\end{equation}
where $\phi$ denotes a charged scalar field. One is thus obviously led to three topologically distinct types of one-loop Feynman diagrams contributing to the considered four-photon process, namely the box, triangle, and the bubble, depicted in \rff{f:scalarg}.
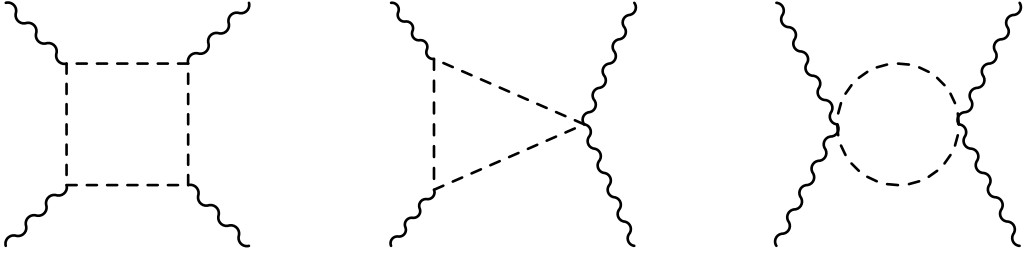
\begin{figure}[ht]
\hspace*{\fill}
\begin{fmfgraph*}(115,92)
\fmfleft{i1,i2}
\fmfright{o1,o2}
\fmf{dashes}{v1,v4}
\fmf{dashes}{v2,v1}
\fmf{dashes}{v3,v2}
\fmf{dashes}{v4,v3}
\fmf{wiggly,tension=2}{v1,i1}
\fmf{wiggly,tension=2}{v2,i2}
\fmf{wiggly,tension=2}{v3,o2}
\fmf{wiggly,tension=2}{v4,o1}
\end{fmfgraph*}
\hfill
\begin{fmfgraph*}(115,92)
\fmfleft{i1,i2}
\fmfright{o1,o2}
\fmf{dashes}{v1,v3}
\fmf{dashes}{v2,v1}
\fmf{dashes}{v3,v2}
\fmf{wiggly,tension=3.5}{v1,i1}
\fmf{wiggly,tension=3.5}{v2,i2}
\fmf{wiggly,tension=3}{v3,o2}
\fmf{wiggly,tension=3}{v3,o1}
\end{fmfgraph*}
\hfill
\begin{fmfgraph*}(115,92)
\fmfleft{i1,i2}
\fmfright{o1,o2}
\fmf{dashes,right}{v1,v2}
\fmf{dashes,right}{v2,v1}
\fmf{wiggly,tension=2}{v1,i1}
\fmf{wiggly,tension=2}{v1,i2}
\fmf{wiggly,tension=2}{v2,o2}
\fmf{wiggly,tension=2}{v2,o1}
\end{fmfgraph*}
\hspace*{\fill}
\caption{Scalar QED: Box, triangle, and bubble diagrams contribute. Dashed lines represent propagators of the charged massive scalar field.}
\label{f:scalarg}
\end{figure}

As in the case of spinor QED, all of them are individually only logarithmically divergent. The UV divergences mutually cancel in the sum of the three diagrams, and the fate of the gauge non-invariant finite terms (involving the wrong number of the external momenta) is the same. Unlike spinor QED, the latter terms drop out only upon using the kinematic identity \rf{e:identity}. The rest of the computation is essentially the same as before, and the final form for the polarization tensor reads
\begin{align}
\Gamma^\mathrm{scal.}_{pppp} &= N_\mathrm{scal.} \sum_\mathrm{perm.} -6(1_\nu 1_\rho 2_\sigma 3_\mu) - 1_\rho 1_\sigma 2_\mu 3_\nu\notag\\
\Gamma^\mathrm{scal.}_{gg} &= N_\mathrm{scal.} \sum_\mathrm{perm.} 4(g_{\mu\nu}g_{\rho\sigma} + g_{\mu\sigma}g_{\nu\rho})\dpr{12}^2\notag\\
&\hphantom{{}= N_\mathrm{scal.} \sum_\mathrm{perm.}} + (6g_{\mu\nu}g_{\rho\sigma} + 3g_{\mu\sigma}g_{\nu\rho})\dpr{12}\dpr{13}\notag\\
\Gamma^\mathrm{scal.}_{ppg} &= N_\mathrm{scal.} \sum_\mathrm{perm.} \Big[ 6(1_\rho 3_\mu g_{\nu\sigma}) - 4(1_\sigma 2_\mu g_{\nu\rho} )\notag\\
&\hphantom{{}= N_\mathrm{scal.} \sum_\mathrm{perm.} \Big[} + 2(1_\rho 3_\nu g_{\mu\sigma} + 1_\rho 3_\sigma g_{\mu\nu})\notag\\
&\hphantom{{}= N_\mathrm{scal.} \sum_\mathrm{perm.} \Big[} - 1_\sigma 3_\nu g_{\mu\rho} - 3_\mu 3_\nu g_{\rho\sigma}\notag\\
&\hphantom{{}= N_\mathrm{scal.} \sum_\mathrm{perm.} \Big[} + 3_\mu 3_\sigma g_{\nu\rho}- 3(1_\sigma 3_\mu g_{\nu\rho}) \Big]\dpr{12},
\label{e:scalart}
\end{align}
where $N_\mathrm{scal.} = \alpha^2/45 m^4$, with $m$ being the charged scalar boson mass.

Employing the tricks mentioned in the preceding section (i.e. utilizing appropriately the identity \rf{e:identity}), the expression \rf{e:scalart} can be recast in the desired form
\begin{align}
\Gamma^\mathrm{scal.}_{pppp} &= N_\mathrm{scal.} \sum_\mathrm{perm.} -6(1_\nu 1_\rho 2_\sigma 3_\mu) - 1_\rho 1_\sigma 2_\mu 3_\nu\notag\\
\Gamma^\mathrm{scal.}_{gg} &= N_\mathrm{scal.} \sum_\mathrm{perm.} \bb{\frac{5}{2}(g_{\mu\rho}g_{\nu\sigma}) + g_{\mu\sigma}g_{\nu\rho}}\dpr{12}^2\notag\\
\Gamma^\mathrm{scal.}_{ppg} &= N_\mathrm{scal.} \sum_\mathrm{perm.} \Big[ 1_\sigma 3_\mu g_{\nu\rho} - 1_\sigma 3_\nu g_{\mu\rho} - 3_\mu 3_\nu g_{\rho\sigma} + 3_\mu 3_\sigma g_{\nu\rho}\notag\\
&\hphantom{{}= N_\mathrm{scal.} \sum_\mathrm{perm.} \Big[} +6(1_\rho 3_\mu g_{\nu\sigma} + 1_\rho 3_\sigma g_{\mu\nu})\notag\\
&\hphantom{{}= N_\mathrm{scal.} \sum_\mathrm{perm.} \Big[} +2 (1_\rho 3_\nu g_{\mu\sigma}) \Big]\dpr{12},
\end{align}
and its direct comparison with \rf{e:effect} then yields
\begin{align}
g_1^\mathrm{scal.} = \frac{7 N_\mathrm{scal.}}{N} &= \frac{7\alpha^2}{1440 m^4}\notag\\
g_2^\mathrm{scal.} = \frac{N_\mathrm{scal.}}{N} &= \frac{\alpha^2}{1440 m^4}.
\end{align}

\section{Vector QED}\label{s:vect}
Now we proceed to the technically most difficult case, namely the QED of charged massive vector bosons $W$ (for brevity, we use the term vector QED). As we have noted in \rfs{s:intro}, we will employ the Yang-Mills form of the interaction since it is naturally embedded in the standard electroweak theory formulated in the physical $U$-gauge. The relevant interaction Lagrangian thus can be written as 
\begin{equation}
\mathscr{L}_\mathrm{int.} = \mathscr{L}_{WW\gamma}+\mathscr{L}_{WW\gamma\gamma},
\end{equation}
where
\begin{align}
\mathscr{L}_{WW\gamma}&\equiv -\cu e(A_\mu W_\nu \stackrel{\leftrightarrow}{\partial^\mu} W^{\dagger\nu}+W_\mu W^\dagger_\nu \stackrel{\leftrightarrow}{\partial^\mu} A^\nu+W^\dagger_\mu A_\nu \stackrel{\leftrightarrow}{\partial^\mu} W^\nu)\notag\\
\mathscr{L}_{WW\gamma\gamma}&\equiv -e^2(W_\mu W^{\dagger\mu} A_\nu A^\nu - W_\mu A^\mu W^\dagger_\nu A^\nu).
\end{align}
It is then obvious that the one-loop Feynman diagrams for the considered four-photon process are topologically analogous to those encountered in scalar QED: we have to deal again with the box, triangle, and the bubble (see \rff{f:vectorg}).
\begin{figure}[ht]
\hspace*{\fill}
\begin{fmfgraph*}(115,92)
\fmfleft{i1,i2}
\fmfright{o1,o2}
\fmf{dbl_wiggly}{v1,v4}
\fmf{dbl_wiggly}{v2,v1}
\fmf{dbl_wiggly}{v3,v2}
\fmf{dbl_wiggly}{v4,v3}
\fmf{wiggly,tension=2}{v1,i1}
\fmf{wiggly,tension=2}{v2,i2}
\fmf{wiggly,tension=2}{v3,o2}
\fmf{wiggly,tension=2}{v4,o1}
\end{fmfgraph*}
\hfill
\begin{fmfgraph*}(115,92)
\fmfleft{i1,i2}
\fmfright{o1,o2}
\fmf{dbl_wiggly}{v1,v3}
\fmf{dbl_wiggly}{v2,v1}
\fmf{dbl_wiggly}{v3,v2}
\fmf{wiggly,tension=3.5}{v1,i1}
\fmf{wiggly,tension=3.5}{v2,i2}
\fmf{wiggly,tension=3}{v3,o2}
\fmf{wiggly,tension=3}{v3,o1}
\end{fmfgraph*}
\hfill
\begin{fmfgraph*}(115,92)
\fmfleft{i1,i2}
\fmfright{o1,o2}
\fmf{dbl_wiggly,right}{v1,v2}
\fmf{dbl_wiggly,right}{v2,v1}
\fmf{wiggly,tension=2}{v1,i1}
\fmf{wiggly,tension=2}{v1,i2}
\fmf{wiggly,tension=2}{v2,o2}
\fmf{wiggly,tension=2}{v2,o1}
\end{fmfgraph*}
\hspace*{\fill}
\caption{Vector QED: Box, triangle, and bubble diagrams contribute. The internal lines represent propagators of the charged massive vector field.}
\label{f:vectorg}
\end{figure}
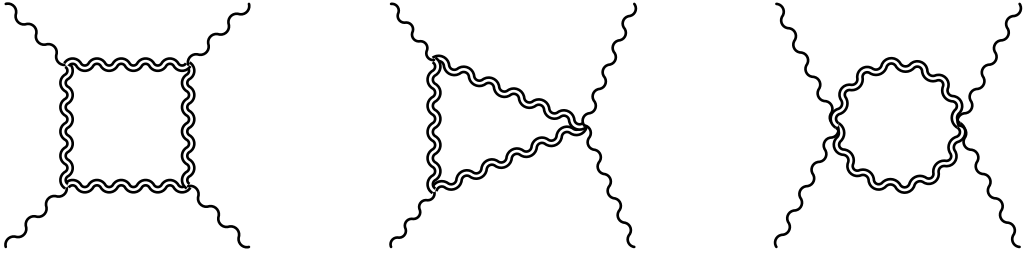

The $W$ boson propagator has the canonical $U$-gauge form 
\begin{equation}
D_{\mu\nu}(k) = \frac{-g_{\mu\nu} + k_\mu k_\nu/m^2}{k^2 - m^2 + \cu \eps},
\label{e:propag}
\end{equation}
and for the reader's convenience, let us also recall that the Feynman rule corresponding to the trilinear interaction $WW\gamma$ involves the function
\begin{equation}
V_{\mu\nu\rho}(k,p,q) = (k-p)_\rho g_{\mu\nu} + (p-q)_\mu g_{\nu\rho}+(q-k)_\nu g_{\mu\rho}.
\label{e:vertex}
\end{equation}
where $k$, $p$, and $q$ denote the four-momenta outgoing from the vertex. It is a common wisdom that the main source of technical difficulties in any $U$-gauge calculations is precisely the form of the propagator \rf{e:propag}: it behaves as a constant at infinity, and this in turn leads to a high degree of UV divergences occurring in the individual diagrams. An extra amount of work is thus needed to eliminate such spurious divergences before arriving at the final meaningful result (in fact, this is the price to be paid for the relatively low number of the relevant diagrams in comparison with the $R$-gauge calculations in SM). In any case, by combining the expressions \rf{e:propag} and \rf{e:vertex} in the contributions of the diagrams in question, one gets a huge number of terms when the necessary algebraic manipulations are worked out in detail (it is easy to guess that the situation is particularly severe for the box diagram), and it would be practically impossible to proceed without using computer algebra systems such as \texttt{Mathematica}.

Let us now summarize briefly the essential steps of our calculational algorithm. In the integral representing a considered diagram, we introduce Feynman parametrization and perform an appropriate shift of the loop momentum $\ell$ so as to make the denominator of the integrand an even function (i.e. a function of $\ell^2$). As a book-keeping device, we will employ the rescaling $\ell\to\xi\ell$ and then expand the numerator of the integrand in powers of the auxiliary parameter $\xi$. This allows us to isolate and process individually the terms involving even powers of $\ell$ (those containing the odd powers drop out immediately upon symmetric integration). The next step is to use the symmetric integration recipe 
\begin{align}
\ell_\mu\ell_\nu &\to \frac{\ell^2}{d}g_{\mu\nu}\notag\\
\ell_\mu\ell_\nu\ell_\varrho\ell_\sigma &\to \frac{\ell^4}{d(d+2)}(g_{\mu\nu}g_{\rho\sigma}+g_{\mu\rho}g_{\nu\sigma}+g_{\mu\sigma}g_{\nu\rho})\notag\\&\ldots,
\label{e:symint}
\end{align}
where $d$ is the spacetime dimension (the parameter of dimensional regularization), and we denote $\ell^{2n} \equiv (\ell^2)^n$. Note that for a product of $L$ loop momenta, the corresponding relation \rf{e:symint} has $(L - 1)!!$ terms on its right-hand side. The rest of the calculation is then essentially the same as in the previous two sections --- it incorporates the low-energy expansion followed by the integration over Feynman parameters.
 
However, there are some special points to be mentioned here. Starting with the box diagram, it is easy to see that here the highest possible divergence would be octic. This corresponds to the power $\ell^{12}$ in the integrand numerator (combined with the power $\ell^8$ in the denominator). It turns out that such a potential divergence vanishes by itself, on purely algebraic grounds. Remarkably, the next-to-leading (sextic) divergence has the same fate. It is worth noting that these cancellations can be observed even before launching the routine of Feynman parametrization and the usual subsequent steps. Concerning quartic and lower divergences, these are eliminated in the sum of the three diagrams in \rff{f:vectorg}, together with all the unwanted finite terms, when dimensional regularization is carried out in the standard way.

Another technical comment is perhaps in order here. With the growing number $L$ of factors in products of the loop momenta, the relations \rf{e:symint} would obviously lead to an excessive proliferation of the terms to be taken into account. Fortunately, it turns out that upon some algebraic manipulations, the maximum relevant value of $L$ is reduced to six (and the corresponding relation \rf{e:symint} thus yields fifteen terms only).

It is also interesting to observe the role played by the kinematic identity \rf{e:identity} in the aforementioned cancellation mechanism in the three considered QED models. In spinor QED, it is not needed at all (neither for the UV divergences nor for the unwanted gauge non-invariant finite terms). In scalar QED, it must be employed for the elimination of the latter. Finally, in vector QED, it is necessary for eliminating both the UV divergences and the unwanted finite terms.

The result of the calculational \emph{tour de force} sketched above then reads 
\begin{align}
\Gamma^\mathrm{vect.}_{pppp} &= N_\mathrm{vect.} \sum_\mathrm{perm.} -36(1_\nu 1_\rho 2_\sigma 3_\mu) - 486(1_\rho 1_\sigma 2_\mu 3_\nu)\notag\\
\Gamma^\mathrm{vect.}_{gg} &= N_\mathrm{vect.} \sum_\mathrm{perm.} (-70g_{\mu\nu}g_{\rho\sigma} + 504g_{\mu\sigma}g_{\nu\rho})\dpr{12}^2\notag\\
&\hphantom{{}= N_\mathrm{vect.} \sum_\mathrm{perm.}} + (328g_{\mu\nu}g_{\rho\sigma} + 18g_{\mu\sigma}g_{\nu\rho})\dpr{12}\dpr{13}\notag\\
\Gamma^\mathrm{vect.}_{ppg} &= N_\mathrm{vect.} \sum_\mathrm{perm.} \Big[
53(1_\sigma 2_\rho g_{\mu\nu} + 1_\nu 3_\sigma g_{\mu\rho}) - 69(1_\nu 1_\sigma g_{\mu\rho} + 1_\rho 1_\sigma g_{\mu\nu})\notag\\
&\hphantom{{}= N_\mathrm{vect.} \sum_\mathrm{perm.} \Big[} - 184(1_\rho 2_\mu g_{\nu\sigma} + 1_\nu 3_\mu g_{\rho\sigma}) - 14(1_\nu 2_\mu g_{\rho\sigma})\notag\\
&\hphantom{{}= N_\mathrm{vect.} \sum_\mathrm{perm.} \Big[} - 108(1_\nu 1_\rho g_{\mu\sigma}) - 557(1_\sigma 2_\mu g_{\nu\rho}) - 540(3_\mu 3_\nu g_{\rho\sigma})\notag\\
&\hphantom{{}= N_\mathrm{vect.} \sum_\mathrm{perm.} \Big[} + 8(1_\rho 3_\mu g_{\nu\sigma}) + 788(1_\rho 3_\nu g_{\mu\sigma}) - 71(1_\sigma 3_\mu g_{\nu\rho})\notag\\
&\hphantom{{}= N_\mathrm{vect.} \sum_\mathrm{perm.} \Big[}- 433(1_\sigma 3_\nu g_{\mu\rho}) + 417(3_\mu 3_\sigma g_{\nu\rho}) - 521(1_\rho 3_\sigma g_{\mu\nu}) \Big]\dpr{12},
\end{align}
where $N_\mathrm{vect.} = \alpha^2/90 m^4$, with $m$ being the vector boson mass. Utilizing the tricks explained earlier, this can be easily recast as
\begin{align}
\Gamma^\mathrm{vect.}_{pppp} &= N_\mathrm{vect.} \sum_\mathrm{perm.} -36(1_\nu 1_\rho 2_\sigma 3_\mu) - 486(1_\rho 1_\sigma 2_\mu 3_\nu)\notag\\
\Gamma^\mathrm{vect.}_{gg} &= N_\mathrm{vect.} \sum_\mathrm{perm.} (-225g_{\mu\nu}g_{\rho\sigma} + 486g_{\mu\sigma}g_{\nu\rho})\dpr{12}^2\notag\\
\Gamma^\mathrm{vect.}_{ppg} &= N_\mathrm{vect.} \sum_\mathrm{perm.} \Big[ 486 (1_\sigma 3_\mu g_{\nu\rho} - 1_\sigma 3_\nu g_{\mu\rho} - 3_\mu 3_\nu g_{\rho\sigma} + 3_\mu 3_\sigma g_{\nu\rho})\notag\\
&\hphantom{{}= N_\mathrm{vect.} \sum_\mathrm{perm.} \Big[} + 36 (1_\rho 3_\mu g_{\nu\sigma} + 1_\rho 3_\sigma g_{\mu\nu})\notag\\
&\hphantom{{}= N_\mathrm{vect.} \sum_\mathrm{perm.} \Big[} +972 (1_\rho 3_\nu g_{\mu\sigma}) \Big]\dpr{12},
\end{align}
which matches correctly the tensor structure in \rf{e:effect}. The relevant effective coupling constants are then
\begin{align}
g_1^\mathrm{vect.} = \frac{522 N_\mathrm{vect.}}{N} &= \frac{29\alpha^2}{160 m^4}\notag\\
g_2^\mathrm{vect.} = \frac{486N_\mathrm{vect.}}{N} &= \frac{27\alpha^2}{160 m^4}.
\end{align}

\section{Conclusion}\label{s:concl}
The main goal of the present work is a direct diagrammatic evaluation of the effective Lagrangian of the Euler-Heisenberg type for light-by-light scattering in the electrodynamics of charged massive vector bosons. This QED model is naturally embedded in the standard electroweak theory, and we have performed the calculation at the one-loop level in the physical $U$-gauge. We believe that such a calculation has not been published before since the $R$-gauge formalism is usually adopted as more feasible for standard model loop calculations. We have confirmed the old result \cite{vanyashin} obtained by means of a functional technique.

Our diagrammatic calculation may be viewed as a rather laborious technical exercise, but it is certainly gratifying that we have been able to reproduce the result \cite{vanyashin} by means of an entirely different method. For the sake of completeness, we have also recovered the corresponding (perhaps better known) result for scalar QED. For validation of our methods and for comparison with other models, we have of course discussed first the familiar reference case of spinor QED. Our results are neatly summarized in \rft{t:consts}, where we display the ``reduced" effective coupling constants $\tilde{g}_1$ and $\tilde{g}_2$ defined in terms of the original constants $g_1$ and $g_2$ appearing in \rf{e:lagr} as $g_{1,2}\equiv\tilde{g}_{1,2} \alpha^2/m^4$. Let us remark that our results are also consistent with those shown in the recent papers \cite{fichet2014} and \cite{baldenegro}, where the heat kernel method has been employed for the calculation of the relevant effective Lagrangians.

\begin{table}[ht]
\begin{center}
\begin{tabular}{|l|r|r|}
\hline
\textbf{Version of QED} & $\tilde{g}_1$ & $\tilde{g}_2$ \\
\hline
spinor & $1/90$ & $7/360$\\
scalar & $7/1440$& $1/1440$\\
vector & $29/160$ & $27/160$\\
\hline
\end{tabular}
\end{center}
\caption{Reduced effective coupling constants for the considered QED models.}
\label{t:consts}
\end{table}
Light-by-light scattering is a fundamental quantum process, which should exist beyond any reasonable doubt, and its characteristics have been theoretically predicted since the early days of modern QED. It is also well known that its direct detection is an extremely difficult experimental task, and thus it remained elusive for many years.

However, quite recently, there has been a remarkable progress in this direction, maybe from a somewhat unexpected side: it turns out that it is possible to detect the scattering of (quasi)real photons produced by colliding Pb-Pb beams at the LHC (see \cite{aaboud} for the original report of the measurements performed by the ATLAS Collaboration). Phenomenological analysis of such a setup can be found in several recent papers, see e.g. \cite{denterria}, \cite{szczurek}, \cite{fichet2016}, and \cite{ellis}. Thus, it is encouraging to see that such an old topic related to the basics of quantum field theory still retains its interest and exhibits some potential for a further research.

\section*{Acknowledgments}
One of us (FP) thanks Dr. Karol Kampf and Dr. Ji\v{r}\'{i} Novotn\'{y} for their help with \texttt{Mathematica} and for useful discussions. This work was supported by the Czech Science Foundation grant No. GACR15-180805.
\end{fmffile}

\end{document}